\documentclass[aps,
prd,twocolumn,eqsecnum,nofootinbib,floatfix]{revtex4-1}

\usepackage{times}
\usepackage{xspace} % important for correct spacing - see command list

\usepackage{chngcntr}
\usepackage{graphicx} 
\usepackage{bm}
\usepackage{amssymb}  
\usepackage[pdftex]{color}
\usepackage{slashed}
\usepackage{booktabs}
\usepackage{multirow}
\usepackage{mathtools}
\usepackage{amsmath}
\usepackage{stackrel}
\usepackage{rotating}
\usepackage{rotfloat}
\usepackage{hyperref}
\hypersetup{
    colorlinks=true,       % false: boxed links; true: colored links
    linkcolor=blue,          % color of internal links
    citecolor=blue,        % color of links to bibliography
    filecolor=blue,      % color of file links
    urlcolor=blue           % color of external links
}
\usepackage{amsmath}
\usepackage[utf8]{inputenc}
\usepackage{cleveref}

\definecolor{red}{rgb}{0.8,0.0,0.0}
\definecolor{green}{rgb}{0.0,0.6,0.0}
\definecolor{darkblue}{rgb}{0.0,0.1,0.7}
\definecolor{brown}{rgb}{0.6,0.1,0.0}
\definecolor{grey}{rgb}{0.6,0.6,0.6}
\definecolor{darkgreen}{rgb}{0.0, 0.545098, 0.0}
\definecolor{applegreen}{rgb}{0.55, 0.71, 0.0}
\definecolor{purple}{rgb}{0.5,0.0,0.5}
\definecolor{babypink} {rgb}{0.64, 0.44, 0.44}
\definecolor{orange}{rgb}{1.0,0.5,0.0}

\definecolor{DARKBLUE}{rgb}{0.0,0.1,0.7}

\allowdisplaybreaks

\newcommand{\ga}{\gamma}

\newcommand{\bi}{\begin{itemize}}
\newcommand{\ei}{\end{itemize}}

\newcommand{\ben}{\begin{enumerate}}
\newcommand{\een}{\end{enumerate}} 

\newcommand{\bt}[1]{\begin{table}[tb]\begin{tabular}{#1} \hline\hline  \\[-1.0em]}
\newcommand{\et}[2]{\hline\hline \end{tabular} \caption{#1} \label{#2} \end{table}}

\newcommand{\be}{\begin{equation}}
\newcommand{\ee}{\end{equation}}

\newcommand{\bea}{\begin{eqnarray}}
\newcommand{\eea}{\end{eqnarray}}

\renewcommand{\Im}{\ensuremath{\mathop{\mathrm{Im}}}}

\renewcommand{\case}[2]{\ensuremath{{\textstyle\frac{#1}{#2}}}}
\newcommand{\half}{\ensuremath{\case{1}{2}}}

\newcommand{\gaga}{\gamma\gamma}
\newcommand{\gagagaga}{\ensuremath{\gamma\gamma\rightarrow\gamma\gamma}\xspace}

\newcommand{\mev}{\ensuremath{\mathrm{\,Me\kern -0.1em V}}\xspace}
\newcommand{\gev}{\ensuremath{\mathrm{\,Ge\kern -0.1em V}}\xspace}

\begin{document}

\title{The two-photon decay of  X(6900) from  light-by-light scattering at the LHC}
\author{Volodymyr Biloshytskyi}
\author{Vladimir Pascalutsa}
\affiliation{Institut f\"ur Kernphysik,
 Johannes Gutenberg-Universit\"at  Mainz,  D-55128 Mainz, Germany}
\author{Lucian Harland-Lang}
\affiliation{Rudolf Peierls Centre, Beecroft Building, Parks Road, Oxford, OX1 3PU}
\author{Bogdan Malaescu}
\affiliation{LPNHE, Sorbonne Universit\'e, Universit\'e Paris Cit\'e, CNRS/IN2P3, Paris, France}
\author{Kristof Schmieden}
\author{Matthias Schott}
\affiliation{Institut f\"ur Physik,
 Johannes Gutenberg-Universit\"at  Mainz,  D-55128 Mainz, Germany}

\date{\today}

\begin{abstract}

The LHCb Collaboration has recently discovered a structure around 6.9 GeV in the double-$J/\psi$ mass distribution, possibly a first fully-charmed tetraquark state $X(6900)$. Based on vector-meson dominance (VMD) such a state should have a significant branching ratio for decaying into two photons.
We show that the recorded LHC data for the light-by-light scattering may indeed accommodate for such a state, with a $\gamma \gamma$ branching ratio of order of $10^{-4}$, which is larger  even than the value inferred by the VMD. The spin-parity assignment $0^{-+}$ is in better agreement with the VMD prediction than $0^{++}$, albeit not significantly at the current precision. Further light-by-light scattering data in this region, clarifying the nature of this state, should be obtained in the Run 3 and probably in the high-luminosity phase of the LHC (Run 4 etc.).
\end{abstract}

\maketitle
\counterwithout{equation}{section}
\section{Introduction}

The ATLAS and CMS Collaborations have recently made first  experimental observations of
 light-by-light (LbL) scattering  in the ultra-peripheral Pb-Pb collisions at the LHC \cite{ATLAS2017,CMS2019}. The ATLAS Collaboration
has subsequently provided the most comprehensive dataset from the LHC Run-2 \cite{ATLAS2020}, which shows a mild excess over the Standard Model prediction
centered on the diphoton invariant mass region of 5 to 10 GeV
(cf.\ Fig.~\ref{fig:MainFigure} below). A similar excess between 5-7 GeV of the diphoton invariant mass was seen by CMS Collaboration \cite{CMS2019} as well.

More recently, the LHCb Collaboration
has observed a structure in the di-$J/\psi$ mass distribution \cite{LHCb2020} and interpreted it as a new state,  $X(6900)$, with mass and di-$J/\psi$ width
quoted in Table~\ref{tab:tetraquark}.
This state is possibly the lightest fully-charmed tetraquark state \cite{Richard2020,Sonnenschein2020,Faustov2021,Deng2020,Guo2020} (see also \cite{Chen2022} for review), and according to Refs.~\cite{Karliner2020,Debastiani2019,Liu2020,Lu2020,Wu2016,Bedolla2019,Wang2019,Wan2020,Liang2021,Li2021,Ke2021} can be a pseudoscalar $P$-wave state ($J^{PC}=0^{-+}$), or a scalar $S$-wave  state ($J^{PC}=0^{++}$). 
A possibility for it to be a tensor meson ($J^{PC}=2^{++}$) is discussed in \cite{Chen2022, Faustov2021,Deng2020,Karliner2020,Lu2020,Bedolla2019,Liang2021,Li2021,Ke2021, Weng2020,Zhu2020}.
In any of these cases, this state would likely couple to two photons and hence contribute to the LbL scattering. In fact, the vector-meson dominance (VMD) hypothesis provides a rather accurate prediction for the two-photon decay width ($X\to\gaga$) in terms of the di-$J/\psi$ width (cf.\ Appendix).

\begin{table}[h]
\centering
\begin{tabular}{c|c|c}
\hline\hline
    Parameter & Interference & No-interference\\
    \hline\hline
    $m_X$ [MeV] & $6886\pm11\pm11$ & $6905\pm11\pm7$  \\
    $\Gamma_{X\rightarrow J/\psi J/\psi}$ [MeV] & $168\pm33\pm69$ & $80\pm19\pm33$ \\
    \hline
    $\Gamma_{X\rightarrow \gamma\gamma}$ [keV]  & $67^{+15}_{-19} $ & $45^{+11}_{-14} $ \\
    \hline\hline
\end{tabular}
\caption{\label{tab:tetraquark} The mass and di-$J/\psi$ width of $X(6900)$ in the two scenarios of Ref.~\cite{LHCb2020}, and the
corresponding two-photon widths obtained here by fitting the light-by-light scattering data of Ref.~\cite{ATLAS2020}.}
 \end{table} 
 
In this work
we explore the possibility of the excess seen in ATLAS experiment is due to the $X(6900)$ meson. The two-photon decay width of this state can then be determined from a fit to the data, with the resulting values shown in the last row of Table~\ref{tab:tetraquark}.
In what follows we describe our formalism for the inclusion of mesons
in LbL scattering (Sec.~\ref{sec:Formalism}), the details and results of the fit to 
ATLAS data (Sec.~\ref{sec:Fitresults}), comparison with VMD estimates (Sec.~\ref{sec:VMDcomparison}),
and conclusions (Sec.~\ref{sec:Conclusions}).

\section{Meson exchange in light-by-light scattering}
\label{sec:Formalism}
We start with outlining the formalism for the inclusion of 
meson states into the LbL process. These states ought to be added at the amplitude level. It is conventional to work with helicity 
amplitudes $M_{\lambda_1 \lambda_2 \lambda_3 \lambda_4} (s,t,u)$, where $\lambda_i= \pm 1$ is the helicity of each of the four photons and the Mandelstam variables of the LbL scattering satisfy the kinematic constraint: $s+t+u=0$. Thanks to the discrete ($P$, $T$, $C$) symmetries only 5 of the 16 amplitudes are independent, e.g.:
  $M_{++++}$, $M_{+--+}$, $M_{+-+-}$, $M_{+++-}$ and $M_{++--}$. Furthermore, the crossing symmetry infers the following relation: 
\begin{equation}
    M_{++++}(s,t,u)=M_{+--+}(t,s,u)= M_{+-+-}(u,t,s).
\end{equation}
The remaining two amplitudes are fully crossing invariant.

In what follows we consider spin-0 mesons, with parity $P=+$ (scalars) or $P=-$ (pseudoscalars). Their tree-level contributions to the LbL amplitudes follow from a simple effective Lagrangian (cf.\ Appendix), yielding the following expressions:
\begin{subequations}
\begin{align}
&M^{P}_{++++}(s,t,u) = -
\frac{16\pi s^2\Gamma_{\gaga}}{m^3\,(s-m^2)},\label{M1}\\
&M^{P}_{+++-}(s,t,u) = 0,\label{M2}\\
&M^{P}_{++--}(s,t,u) \nonumber\\
&\quad=
    -P \frac{16\pi\Gamma_{\gaga}}{m}\left(\frac{s}{s-m^2}+ \frac{t}{t-m^2} + \frac{u}{u-m^2}\right),
\label{M3}
\end{align}
\end{subequations}
where $P=\pm 1$ stands for the parity of the state, $m$ for the mass, and $\Gamma_{\gamma\gamma}$ for the two-photon width.

The nonvanishing amplitudes are precisely the ones entering the forward LbL scattering sum rules \cite{Pascalutsa:2010sj}, and
it is useful to check the consistency of the above expressions with the sum rules. We recall that
the helicity amplitudes of the forward ($t=0$) [or, equally, the backward ($u=0$), scattering of real photons satisfy exact sum rules \cite{Pascalutsa:2010sj,Pascalutsa:2018ced}:  
\begin{subequations}
\begin{align}
M_{++++}(s)+M_{+-+-}(s) &= \frac{2s^2}{\pi}\int\limits_{0}^{\infty}ds'\frac{\sigma_{0}(s')+\sigma_{2}(s')}{s^{'2}-s^2-i0^+},\label{sumrule1}\\
M_{++++}(s)-M_{+-+-}(s) &= \frac{2s}{\pi}\int\limits_{0}^{\infty}ds'\,\frac{s'\left[\sigma_{0}(s')-\sigma_{2}(s')\right]}{s^{'2}-s^2-i0^+},\label{sumrule2}\\
M_{++--}(s) &= \frac{2s^2}{\pi}\int\limits_{0}^{\infty}ds^{\prime}\frac{\sigma_{\parallel}(s')-\sigma_{\perp}(s')}{s^{'2}-s^2-i0^+}.\label{sumrule3}
\end{align}
\end{subequations}
where the right-hand side involves integrals of total $\gamma\gamma$-fusion cross sections for various photon polarizations. 
For the case of $\gamma\gamma$-fusion into a scalar or a pseudoscalar meson these cross sections take the following simple form (see, e.g., \cite{Budnev:1975poe,Pascalutsa2012}):
\begin{subequations}
\begin{eqnarray}
&&\sigma_0(s)=16\pi^2\frac{\Gamma_{\gaga}}{m}\,\delta(s-m^2),\quad \sigma_2(s)=0,\\
&&\begin{cases}
    \sigma_{\parallel}(s)=\sigma_0(s),\quad  \sigma_{\perp}(s)=0, \quad \text{for scalar},\\
     \sigma_{\perp}(s)=\sigma_0(s),\quad  \sigma_{\parallel}(s)=0,\quad \text{for pseudoscalar}.
\end{cases}
\end{eqnarray}
\end{subequations}

Substituting these cross sections into the sum rules we find that 
the contribution to $M_{++++}$ found in Eq.~(\ref{M1}) is reproduced by the first sum rule, but not the second one. This inconsistency can be fixed by reducing the one power of $s$ in the expression (\ref{M1}), thus resulting in:
\begin{equation}
    M^{P}_{++++}(s,t,u) = -
\frac{16\pi s\,\Gamma_{\gaga}}{m\,(s-m^2)}.
\end{equation}
This contribution is consistent with both sum rules and has a better energy behavior. We shall use it in place of Eq.~(\ref{M1}).

The contribution to $M_{++--}$ in Eq.~(\ref{M3}) is consistent with the sum rule (\ref{sumrule3}). As a side remark we note that it  satisfies a more general off-forward sum rule: 
\begin{eqnarray}
M_{++--}(s,t,u) & = & \frac{1}{\pi}\int\limits_{0}^{\infty}ds' \big[\sigma_{\parallel}(s')-\sigma_{\perp}(s')\big] \nonumber \\
&\times &    \Big(\frac{s}{s^{\prime}-s}+\frac{t}{s^{\prime}-t}+\frac{u}{s^{\prime}-u}\Big).
\end{eqnarray}
Any single-meson-exchange contribution to this LbL scattering 
should satisfy this sum rule. However it does not hold in a more general case --- a subtraction function must be added.
A similar off-forward sum rule holds for the crossing-invariant combination $M_{++++}+M_{+--+}+M_{+-+-}$ and the unpolarized
cross section of $\gaga$ fusion. It also  holds without subtraction for the single-meson-exchange contributions.

Next step is the inclusion of the decay width. It can be done by resumming 
the meson self-energy, $\Pi(s)$, in $s$-channel exchange contribution, such that the factors $1/(s-m^2)$ in the above expressions are replaced with 
 $1/\big(s-m^2 - \Pi(s)\big)$. The decay width
 then comes from the imaginary part of the self-energy, i.e., $\Im \, \Pi(s) = - \sqrt{s} \, \Gamma(s)$. The real part of the self-energy contributes to the mass and field renormalization; any further effects of the real part are neglected here. For the total decay width of $X(6900)$-meson we use below the energy-dependent di-$J/\psi$ width, as calculated in the Appendix. 
 
\section{Fitting $X(6900)$ into the light-by-light data}
\label{sec:Fitresults}
We have extended the Monte-Carlo code SuperChic v3.05 \cite{HarlandLang2019,HarlandLang2016}\footnote{Although this is not the most recent version, subsequent updates do not relate to LbL scattering.}
used in the original interpretation of the ATLAS data \cite{ATLAS2020}, by including the $X(6900)$ along with the well-known bottomonium states \cite{Wang2018} pertinent to this energy region, see Table~\ref{tab:bottom_include}.    Note that SuperChic v3.05 includes otherwise 
only the simplest perturbative-QCD (pQCD) contributions to LbL scattering, i.e., the quark-loop contribution.  The next-to-leading order pQCD corrections were shown to contribute at the order of few percent \cite{Bern:2001dg,Klusek-Gawenda:2016nuo,Krintiras:2022jxa}, which is negligible at the current level of experimental precision.

\begin{table}[H]
\centering
\begin{tabular}{ c | c | c | c | c }
\hline\hline
    Meson & $J^{PC}$ & $M$, [MeV] &  $\Gamma_{\mathrm{tot}}$, [MeV] & $\Gamma_{\gamma\gamma}/\Gamma_{\mathrm{tot}}$ [\%]\\
    \hline\hline
    $\eta_b$(1S) & $0^{-+}$ & $9399.0$ & $17.9$ & $5.87\times10^{-3}$\\
    $\eta_b$(2S) & $0^{-+}$ & $9999.0$ & $8.34$ & $5.86\times10^{-3}$ \\
    \hline
    $\chi_{b0}$(1P) & $0^{++}$ & $9859.44$ & $3.39$ & $5.87\times10^{-3}$\\
    $\chi_{b0}$(2P) & $0^{++}$ & $10232.5$ & $3.54$ & $5.41\times10^{-3}$ \\
    \hline\hline
\end{tabular}
\caption{\label{tab:bottom_include} Bottomonium resonances included in this work.}
\end{table}

Given the mass and width of $X(6900)$ from the LHCb determination,  the two-photon-decay width can be determined from the ATLAS data on LbL scattering.  
In the narrow resonance approximation, the LbL cross section depends only on the ratio  $\Gamma_{X\to\gaga}/\sqrt{\Gamma_\mathrm{tot}}$, and hence
we take it as a fitting parameter. The total width is assumed to be dominated by the di-$J/\psi$ decay (i.e., $\Gamma_\mathrm{tot}\simeq \Gamma_{X\to J/\psi\, J/\psi}$). 

\begin{figure}[tbh]
    \centering
    \includegraphics[width=0.9\linewidth]{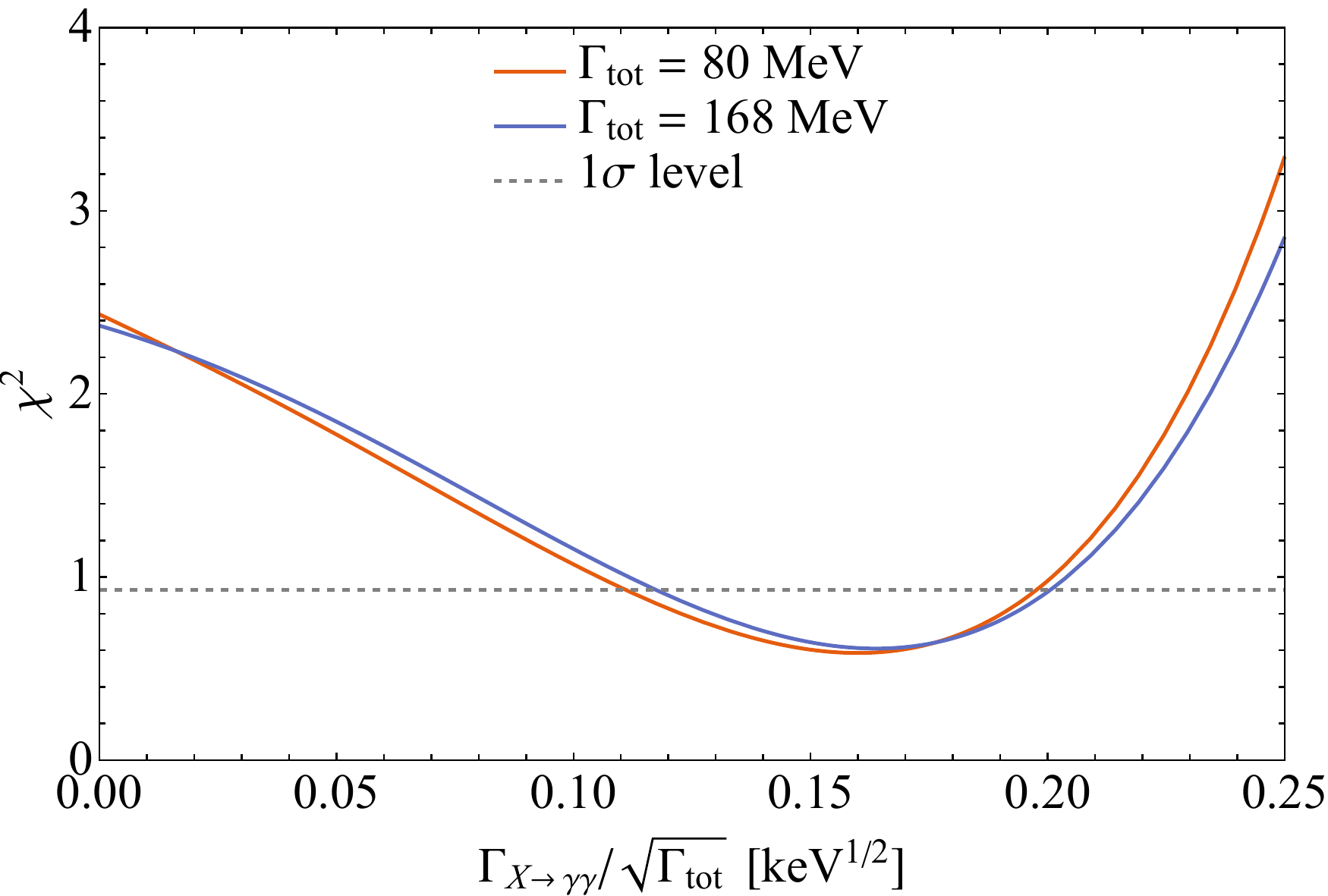}
    \caption{The profile of $\chi^2$ (divided by \#d.o.f. = 3)  for the values of $\Gamma_{\mathrm{tot}}$ used in the two LHCb scenarios. The gray dashed line cuts out the $1\sigma$ interval.}
    \label{fig:chi2slice}
\end{figure}
The fit has been performed to the unfolded diphoton invariant mass spectrum of the ATLAS data. The CMS data is not used  in the present analysis since the corresponding spectrum is not unfolded.
We have explored both the scalar and pseudoscalar nature of $X(6900)$, but the corresponding results of the fit turn out to be  indistinguishable at the current level of statistical accuracy. 
%The tensor possibility ($J^{PC}=2^{++}$) has not been considered, since the results are expected to be similar at the current level of accuracy.
 We therefore show only 
the results for the scalar $X(6900)$.
Since the main uncertainties
in ATLAS data has a statistical origin, then for reasons of
simplicity we take the total experimental uncertainties as the
uncertainties for $\chi^2$ function. The resulting $\chi^2$  is shown in Fig.~\ref{fig:chi2slice}, 
for the two scenarios provided by the LHCb experiment. The best fit yields the following branching ratio 
($\Gamma_{X\to\gamma\gamma}/\Gamma_\mathrm{tot}$):
\begin{align}
&B(X\to\gamma\gamma) =
\begin{cases}
     5.6^{+1.3}_{-1.6}\times 10^{-4},& \text{No-int. sc},\\
     4.0^{+0.9}_{-1.1}\times 10^{-4},& \text{Int. sc.}.
\end{cases}
    \label{Br}
\end{align}
The corresponding values for the $\gaga$ decay width are given in the last row of Table~\ref{tab:tetraquark}. 

\begin{figure*}
    \centering
   \includegraphics[width=0.47\linewidth]{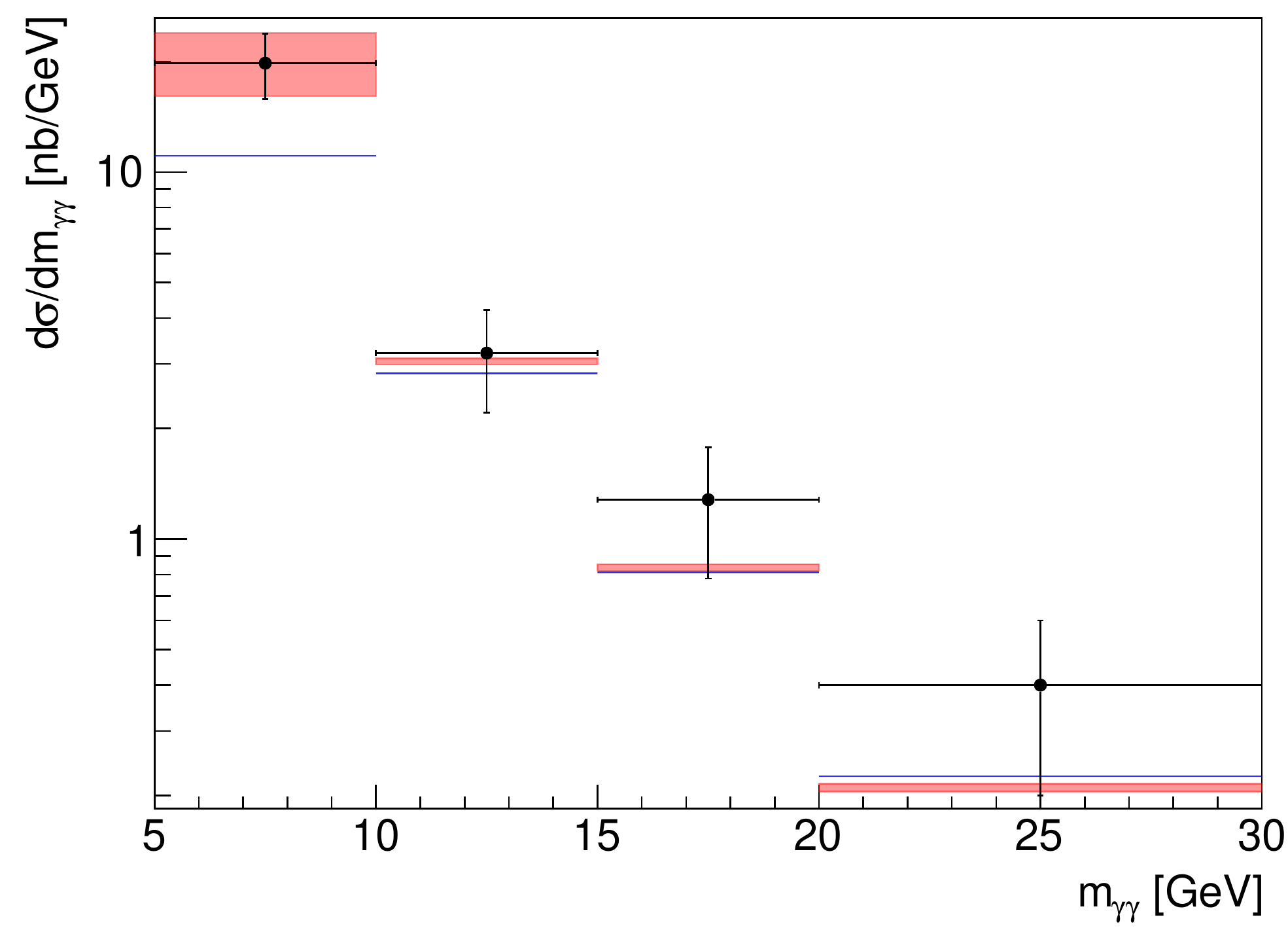}
   \includegraphics[width=0.47\linewidth]{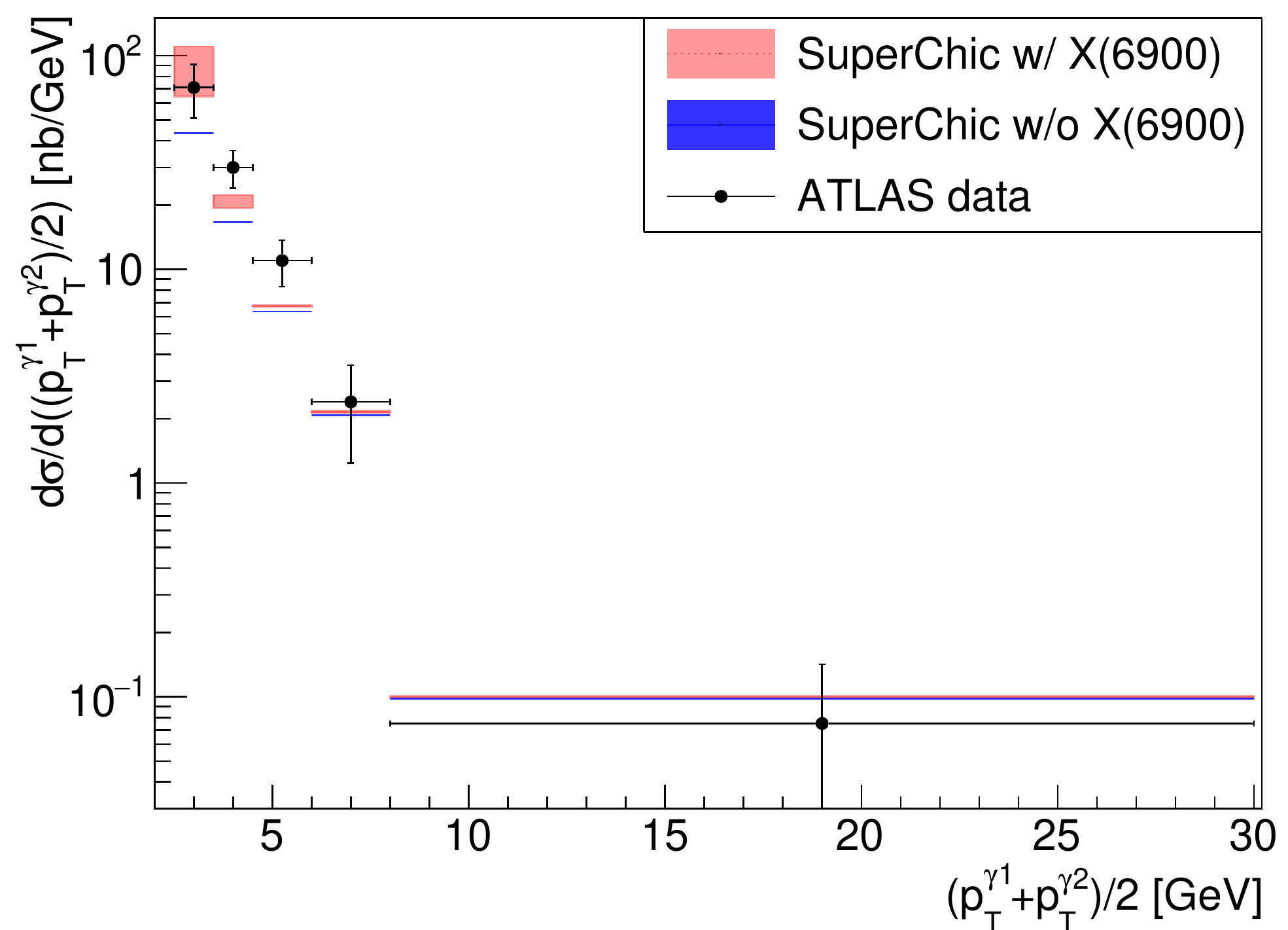}
   \includegraphics[width=0.47\linewidth]{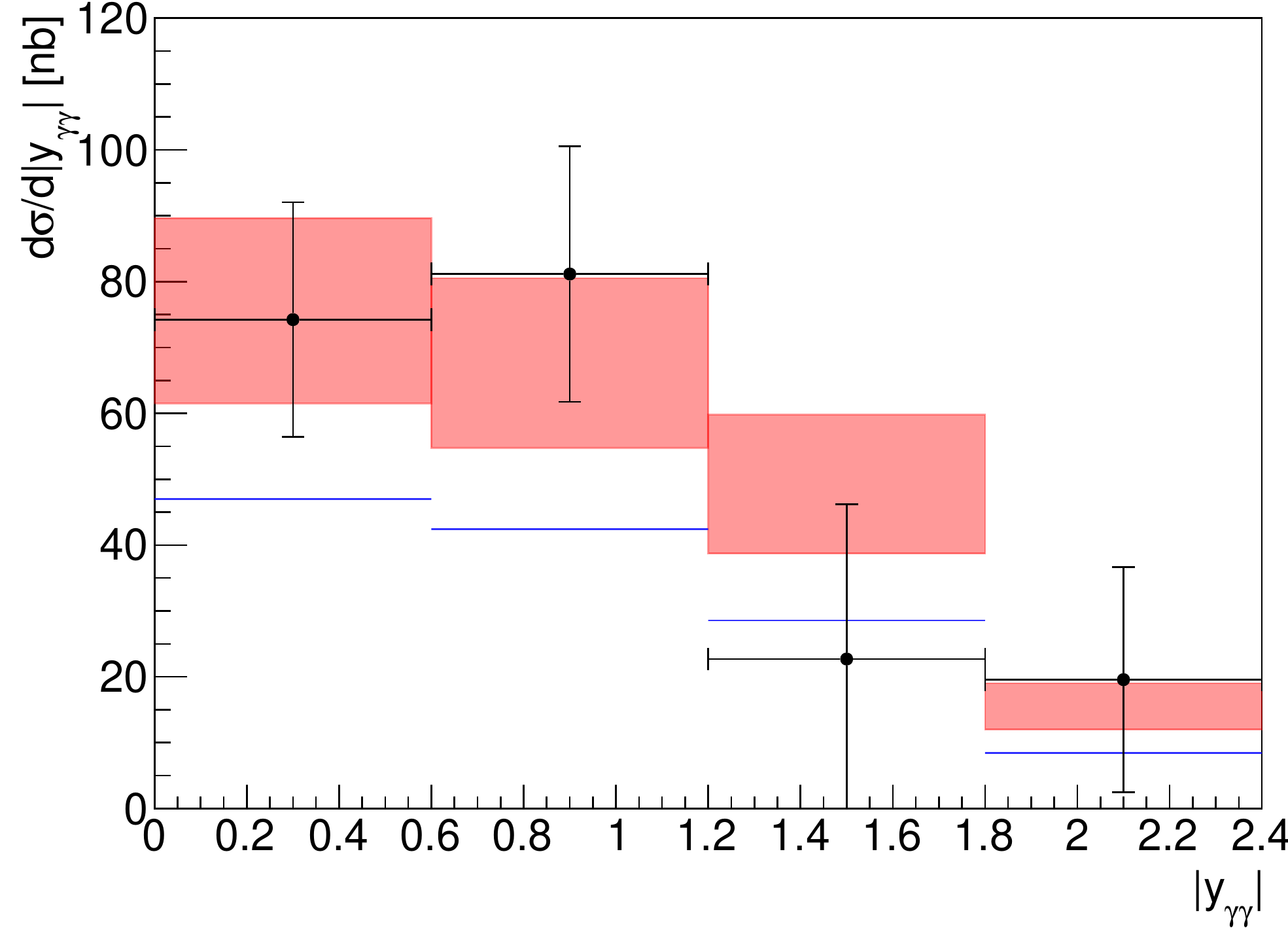}
   \includegraphics[width=0.47\linewidth]{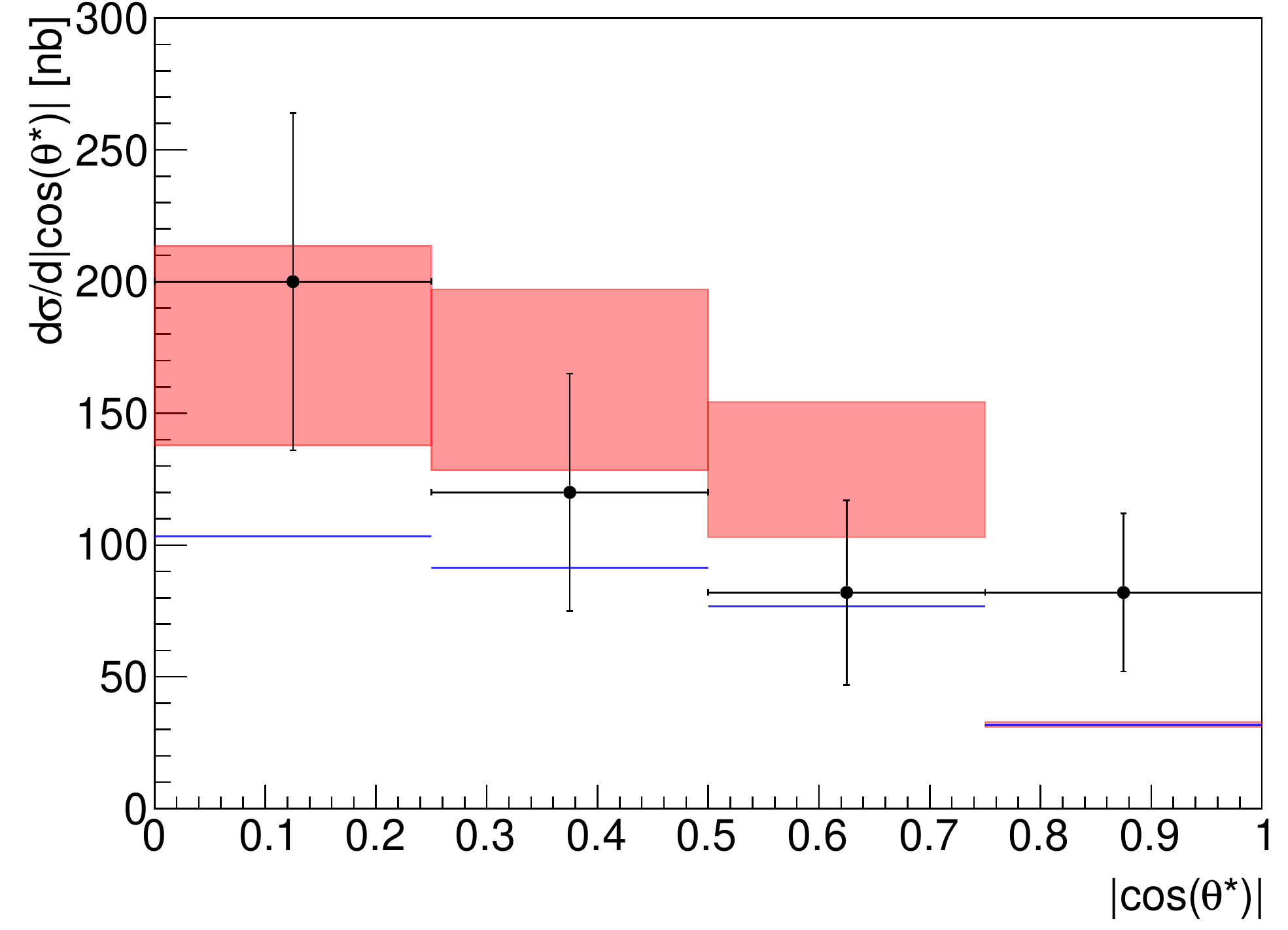}
    \caption{Differential fiducial cross sections of $\gagagaga$ production in Pb$+$Pb collisions at a centre-of-mass energy of $\sqrt{s_{NN}}=5.02$ TeV with integrated luminositiy 2.2 nb$^{-1}$ for four observables (from left to right and top to bottom): diphoton invariant mass $m_{\gaga}$, diphoton absolute rapidity $|y_{\gaga}|$, average photon transverse momentum $(p_{T}^{\ga_1}+p_{T}^{\ga_2})/2$ and diphoton $|cos(\theta^{\ast})|\equiv|\tanh(\Delta y_{\ga_1,\ga_2}/2)|$. The red band represents an uncertainty ($1\sigma$ range) of the fit with $X(6900)$. The blue band contains only the statistical uncertainty of the SuperChic simulation without $X$-resonance.}
    \label{fig:MainFigure}
\end{figure*}

Figure~\ref{fig:MainFigure} shows the exclusive differential cross sections with  and without the inclusion of $X(6900)$, versus the the ATLAS data \cite{ATLAS2020}. The statistical uncertainties of the SuperChic results were highly reduced by simulating a large enough number of events ($10^4$), thus they were neglected in analysis and are not visible on the plots of  Fig.~\ref{fig:MainFigure}. 
The fit yields the integrated fiducial cross section of $\sigma^X_\mathrm{fid}=121\pm20$ nb. It can be compared with the reference SuperChic value without $X$-resonance, $\sigma^0_\mathrm{fid}=76$ nb and with the experimental value, $\sigma^\mathrm{exp.}_\mathrm{fid}= 120 \pm 17 \mathrm{(stat.)} \pm 13 \mathrm{(syst.)} \pm
4 \mathrm{(lumi.)}$ nb, reported by ATLAS \cite{ATLAS2020}.
The description of ATLAS data with $X(6900)$ is better than without it by about $2.3\sigma$.

\section{Vector-meson-dominance estimate}
\label{sec:VMDcomparison}

\begin{figure}[h]
    \centering
   \includegraphics[width=0.4\linewidth]{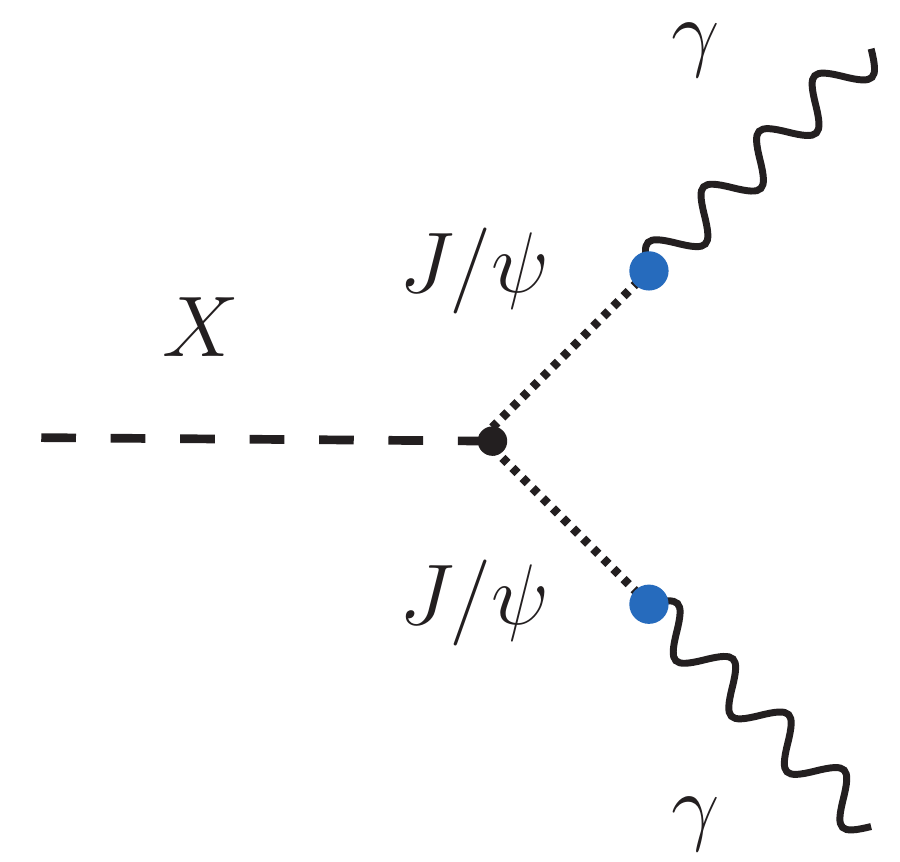}
    \caption{The $X(6900)\to\gaga$ decay via the VMD mechanism.}
    \label{fig:VMD}
\end{figure}

The ratio $\Gamma_{X\to\gaga}/\Gamma_{X\to J/\psi J/\psi}$ can be estimated via the VMD mechanism shown in  Fig.~\ref{fig:VMD}. 
As the result, we obtain the following estimate for branching ratios in the scalar and pseudoscalar case, respectively (cf.~Appendix for details):
\begin{subequations}
\begin{align}
&B^{\mathrm{S}}_{\mathrm{VMD}}(X\to\gamma\gamma)=(2.8\pm 0.4) \times 10^{-6},\label{VMDresultSrel}\\
&B^\mathrm{PS}_\mathrm{VMD}(X\to\gamma\gamma)=(6.4\pm 0.8) \times 10^{-6}.\label{VMDresultPSrel}
\end{align}
\end{subequations}

As one can see, the central values of this estimate is about two orders of magnitude smaller than
we obtained from the fit, Eq.~\eqref{Br}; although, given the large uncertainties, the difference is fairly insignificant. 

Certainly, further measurements of both the di-$J/\psi$ and $\gamma \gamma$ channels are desirable to pin down this possible inconsistency with the VMD expectations.
It could perhaps be explained by other exotic resonances in the diphoton mass region from 5 to 10 GeV, which contribute to the observed excess on the $\gamma \gamma$ channel.  
The broad $X(6900)$ structure has already been proposed to be associated with more than one  tetraquark states \cite{Barnea2006,Berezhnoy2011,Karliner2016,Wang2017,Liu2019,Weng2020,Lundhammar2020}; a second resonance could be located at around  $7.2$ GeV, see e.g.,  \cite{LHCb2020,Sonnenschein2020,Wan2020,Zhu2020,Liang2021}.

\section{Conclusion and Outlook}
\label{sec:Conclusions}

We have shown that the new tetraquark state $X(6900)$, observed by LHCb Collaboartion in the di-$J/\psi$ channel, 
could, in principle, account for the excess in the light-by-light scattering seen in the ATLAS and CMS data, 
The inclusion of $X(6900)$ improves the Standard Model prediction in  the corresponding diphoton mass region
of LbL cross sections. The $X\to \gamma\gamma$ branching ratio has been fitted to the ATLAS data.
The result seen in Eq.~\eqref{Br}, however, exceeds the VMD expectations, albeit statistically the discrepancy is 
not severe. Further measurements of the LbL scattering in the 5 to 10 GeV diphoton-mass range are very desirable
to improve the precision.

Going to lower diphoton masses and increasing the statistics of the $\gagagaga$ events in future runs of the LHC will crucially improve the precision of the fit and hence further constrain the properties of the $X(6900)$. Moreover, the prospective double-differential (or even triple-differential) measurements of a pair (or triplet) of the observables depicted in Fig.~\ref{fig:MainFigure}, which show complementary sensitivity to $X$-state, may provide an additional improvement.
Furthermore, since the Landau-Yang theorem forbids the exchange of the spin-1 $X$-resonance, the analysis of real  $\gagagaga$ scattering ought to reduce the amount of possible quantum numbers of $X(6900)$, which are considered in several analyses (see, e.g. \cite{Liu2019,Liu2020}).
Future measurements at LHCb that will allow the partial-wave analysis, could also narrow down the set of possible quantum number configurations.

\counterwithin{equation}{section}

\section*{Acknowledgements}

This work was supported by the Deutsche Forschungsgemeinschaft (DFG) through the Research Unit FOR5327 [Photon-photon interactions in the Standard Model and beyond]. L.H.L. thanks the Science and Technology Facilities Council (STFC) for support via grant awards ST/L000377/1 and ST/T000864/1.

\appendix

\section{Branching ratio estimate using VMD}
\label{VMDapproach}
We estimate the two-photon decay width of $X(6900)$ by exploiting the vector meson dominance (VMD) hypothesis
as shown in Fig.~\ref{fig:VMD}. The VMD implies that the photon $\lvert\gamma\rangle$ couples via a vector-meson state $\lvert V\rangle$ as follows \cite{Barger1975,Redlich2000}:
\begin{equation}
    \lvert\gamma\rangle\to\frac{e}{M_V}f_V\lvert V\rangle,
\end{equation}
where $e$ is the electron charge, $f_V$ is the corresponding vector-meson decay constant, which is observed in $V\to e^+e^-$ decay, and $M_V$ is its mass. 

\begin{figure}[h]
    \centering
   \includegraphics[width=0.6\linewidth]{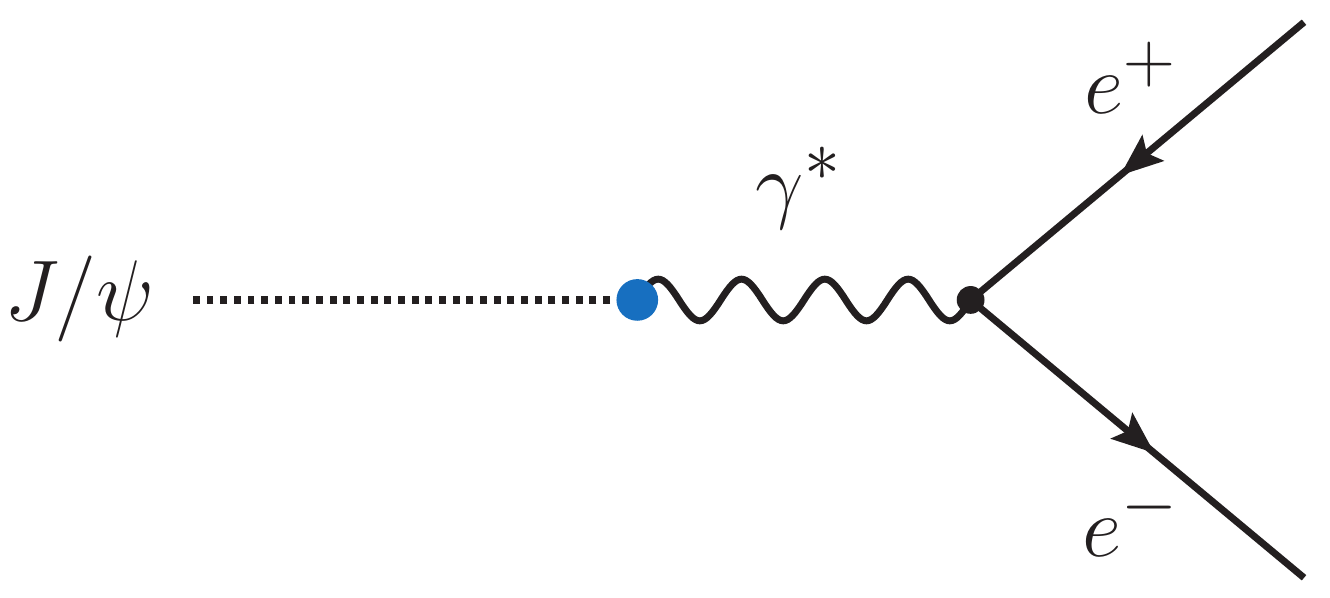}
    \caption{The VMD mechanism of $J/\psi\to e^{+}e^{-}$ decay which determines the $\ga$-$J/\psi$ coupling.}
    \label{fig:Jpsi_ee}
\end{figure}
The $J/\psi$ decay decay constant $f_{\psi}$ can be obtained from the $J/\psi \to e^+e^-$ decay width, cf.~Fig.~\ref{fig:Jpsi_ee}:
\be
\Gamma_{J/\psi\to e^+e^-} =\frac{4\pi\alpha^2f^2_{\psi}}{3 m_{\psi}}.
\ee

Using recent values \cite{PDG2020} for $J/\psi$ mass $m_{\psi}=3096.900\pm0.006$ MeV and electron-positron decay width $\Gamma_{J/\psi\to e^+e^-}=5.55\pm0.17$ keV, one finds $f_{\psi} = 278 \pm 9$ MeV. 

The decay widths $\Gamma_{X\to\gamma\gamma}$ and $\Gamma_{X\to J/\psi J/\psi}$ can be obtained via the imaginary part of the $X$-resonance self-energy derived from the following effective interactions
\begin{align}
    &\mathcal{L}_{X\gamma\gamma} =- g_{X\gaga}\phi_X F^{\mu\nu}F_{\mu\nu},\label{Leff1}\\
    &\mathcal{L}_{X\,J/\psi\gamma} =- 
    g_{X\ga \psi}\phi_X G^{\mu\nu}F_{\mu\nu},\label{Leff2}\\
    &\mathcal{L}_{X\,J/\psi J/\psi} =- 
    g_{X \psi \psi}\phi_X G^{\mu\nu} G_{\mu\nu},\label{Leff3}
\end{align}
where $g_{X\ga\ga}$, $g_{X\ga \psi}$ and $g_{X \psi \psi}$ are dimensionful coupling constants, $F_{\mu\nu} = \partial_{\mu}A_{\nu}-\partial_{\nu}A_{\mu}$ is the photon field tensor, and $ G_{\mu\nu} = \partial_{\mu}B_{\nu}-\partial_{\nu}B_{\mu}$ is the $J/\psi$ field tensor, $\phi_X$ is the scalar field of the $X$-meson. 
Note that we require gauge-invariance with respect to vector fields, including the massive one.
This is where we differ from the recent VMD estimates of Ref.~\cite{Esposito2021}, which begin from a non-invariant
Lagrangian for $J/\psi$.
 For the pseudoscalar case, one of the field tensors is replaced by its dual, i.e., 
\begin{align}
    F^{\mu\nu}&\to\widetilde{F}^{\mu\nu}=\half \epsilon^{\mu\nu\alpha\beta}F_{\alpha\beta},\\
     G^{\mu\nu}&\to\widetilde{G}^{\mu\nu}=\half \epsilon^{\mu\nu\alpha\beta} G_{\alpha\beta}.
\end{align}

The X-V-V vertex that correspond to each of the Lagrangians \eqref{Leff1}-\eqref{Leff3} is
\begin{equation}
    V^{\mu\nu}(q_1,q_2) = -2ig_{XV_1V_2}(q_1\cdot q_2\, g^{\mu\nu}-q_1^{\nu}q_2^{\mu}).
    \label{vertex}
\end{equation}
For the pseudoscalar, the vertex reads as:
\begin{equation}
    \widetilde V^{\mu\nu}(q_1,q_2) = 2i\widetilde g_{XV_1V_2}\epsilon^{\mu\nu\alpha\beta}q_{1\alpha}q_{2\beta}.
    \label{PSvertex}
\end{equation}
Employing the optical theorem, one can write the imaginary part of the self-energy
\begin{equation}
    \Im \Pi_{V_1V_2}(s)= \frac{\lambda^{1/2}(s,m_1^2,m_2^2)}{16\pi s}\sum_{\lambda_1\lambda_2}\big|\mathcal{M}^{\lambda_1\lambda_2}_{X\to V_1V_2}\big|^2,
\end{equation}
where $\lambda_i$ are the helicities, 
\begin{equation}
    \sum_{\lambda_1\lambda_2}|\mathcal{M}^{\lambda_1\lambda_2}_{X\to V_1V_2}|^2
    =\begin{cases} 
    4g^2_{XV_1V_2}\left[2(q_1\cdot q_2)^2+ q_1^2q_2^2\right],\\
    8\widetilde g^2_{XV_1V_2}\left[(q_1\cdot q_2)^2- q_1^2q_2^2\right],
    \end{cases}
\end{equation}
and $\lambda(s,m_1^2,m_2^2)=[s-(m_1+m_2)^2][s-(m_1-m_2)^2]$.
Hence, for the scalar and pseudoscalar cases of $X$(6900) we obtain, 
\begin{align}
    &\Im \Pi_{\gamma\gamma}(s) = \frac{s^2}{16\pi}\theta\left(s\right)\times
    \begin{dcases}
    g^2_{X\gaga},\\
    \widetilde g^2_{X\gaga},
    \end{dcases}\label{SEgg}\\
    &\Im \Pi_{\gamma J/\psi}(s) = \frac{(s-m_{\psi}^2)^3}{8\pi s}\theta\left(s-m^2_\psi\right) \times
    \begin{dcases}
    g^2_{X\gamma \psi} , \\
  \widetilde g^2_{X\gamma \psi},
    \end{dcases}\label{SEgJpsi}
\end{align}
\begin{widetext}
\begin{equation}
    \Im \Pi_{J/\psi J/\psi}(s)= \frac{1}{16\pi}\sqrt{1-\frac{4m_{\psi}^2}{s}} \theta\left(s-4m^2_\psi\right)\times
    \begin{dcases} g^2_{X\psi\psi}
    \big[\left(s-2m_{\psi}^2\right)^2+2m_{\psi}^4\big],\\
     \widetilde g^2_{X\psi\psi} s \big(s-4m_{\psi}^2\big).
    \end{dcases}\label{SEJpsiJpsi}
\end{equation}
Assuming $\Gamma_\mathrm{tot}=\Gamma_{X\to J/\psi J/\psi}$, one thus obtains the following relations between the decay widths of $X(6900)$ into the $\gaga$ and di-$J/\psi$ channels:
\begin{eqnarray}
\Gamma_{X\to\gamma\gamma}^S &=&
\Gamma_{X\to J/\psi J/\psi}\left(\frac{e f_{\psi}}{m_{\psi}}\right)^4\left\{\sqrt{1-\frac{4m_{\psi}^2}{m_X^2}}\left[\left(1-\frac{2m_{\psi}^2}{m_X^2}\right)^2+2\left(\frac{m_{\psi}}{m_X}\right)^4\right]\right\}^{-1},
    \label{GGconnectionS} \\
\Gamma_{X\to\gamma\gamma}^{PS}&=&\Gamma_{X\to J/\psi J/\psi}\left(\frac{e f_{\psi}}{m_{\psi}}\right)^4\left(1-\frac{4m_{\psi}^2}{m_X^2}\right)^{-\frac{3}{2}}.
\label{GGconnectionPS}
\end{eqnarray}
\end{widetext}
Applying these relations, we arrive at the estimate of the branching ratios given in Eqs.~ \eqref{VMDresultSrel} and \eqref{VMDresultPSrel} with corresponding uncertainties that originate from the parameters entering Eqs.\ \eqref{GGconnectionS} and \eqref{GGconnectionPS}, i.e. the $X$(6900) and $J/\psi$ masses and $J/\psi$ decay constant.
\bibliography{./Bib/LbL_LHC}

\begin{thebibliography}{43}%
\makeatletter
\providecommand \@ifxundefined [1]{%
 \@ifx{#1\undefined}
}%
\providecommand \@ifnum [1]{%
 \ifnum #1\expandafter \@firstoftwo
 \else \expandafter \@secondoftwo
 \fi
}%
\providecommand \@ifx [1]{%
 \ifx #1\expandafter \@firstoftwo
 \else \expandafter \@secondoftwo
 \fi
}%
\providecommand \natexlab [1]{#1}%
\providecommand \enquote  [1]{``#1''}%
\providecommand \bibnamefont  [1]{#1}%
\providecommand \bibfnamefont [1]{#1}%
\providecommand \citenamefont [1]{#1}%
\providecommand \href@noop [0]{\@secondoftwo}%
\providecommand \href [0]{\begingroup \@sanitize@url \@href}%
\providecommand \@href[1]{\@@startlink{#1}\@@href}%
\providecommand \@@href[1]{\endgroup#1\@@endlink}%
\providecommand \@sanitize@url [0]{\catcode `\\12\catcode `\$12\catcode
  `\&12\catcode `\#12\catcode `\^12\catcode `\_12\catcode `\%12\relax}%
\providecommand \@@startlink[1]{}%
\providecommand \@@endlink[0]{}%
\providecommand \url  [0]{\begingroup\@sanitize@url \@url }%
\providecommand \@url [1]{\endgroup\@href {#1}{\urlprefix }}%
\providecommand \urlprefix  [0]{URL }%
\providecommand \Eprint [0]{\href }%
\providecommand \doibase [0]{http://dx.doi.org/}%
\providecommand \selectlanguage [0]{\@gobble}%
\providecommand \bibinfo  [0]{\@secondoftwo}%
\providecommand \bibfield  [0]{\@secondoftwo}%
\providecommand \translation [1]{[#1]}%
\providecommand \BibitemOpen [0]{}%
\providecommand \bibitemStop [0]{}%
\providecommand \bibitemNoStop [0]{.\EOS\space}%
\providecommand \EOS [0]{\spacefactor3000\relax}%
\providecommand \BibitemShut  [1]{\csname bibitem#1\endcsname}%
\let\auto@bib@innerbib\@empty
%</preamble>
\bibitem [{\citenamefont {Aaboud}\ \emph {et~al.}(2017)\citenamefont {Aaboud}
  \emph {et~al.}}]{ATLAS2017}%
  \BibitemOpen
  \bibfield  {author} {\bibinfo {author} {\bibfnamefont {M.}~\bibnamefont
  {Aaboud}} \emph {et~al.} (\bibinfo {collaboration} {ATLAS}),\ }\href
  {\doibase 10.1038/nphys4208} {\bibfield  {journal} {\bibinfo  {journal}
  {Nature Phys.}\ }\textbf {\bibinfo {volume} {13}},\ \bibinfo {pages} {852}
  (\bibinfo {year} {2017})}\BibitemShut {NoStop}%
\bibitem [{\citenamefont {Sirunyan}\ \emph {et~al.}(2019)\citenamefont
  {Sirunyan} \emph {et~al.}}]{CMS2019}%
  \BibitemOpen
  \bibfield  {author} {\bibinfo {author} {\bibfnamefont {A.~M.}\ \bibnamefont
  {Sirunyan}} \emph {et~al.} (\bibinfo {collaboration} {CMS}),\ }\href
  {\doibase 10.1016/j.physletb.2019.134826} {\bibfield  {journal} {\bibinfo
  {journal} {Phys. Lett. B}\ }\textbf {\bibinfo {volume} {797}},\ \bibinfo
  {pages} {134826} (\bibinfo {year} {2019})}\BibitemShut {NoStop}%
\bibitem [{\citenamefont {Aad}\ \emph {et~al.}(2021)\citenamefont {Aad} \emph
  {et~al.}}]{ATLAS2020}%
  \BibitemOpen
  \bibfield  {author} {\bibinfo {author} {\bibfnamefont {G.}~\bibnamefont
  {Aad}} \emph {et~al.} (\bibinfo {collaboration} {ATLAS}),\ }\href {\doibase
  10.1007/JHEP03(2021)243} {\bibfield  {journal} {\bibinfo  {journal} {JHEP}\
  }\textbf {\bibinfo {volume} {03}},\ \bibinfo {pages} {243} (\bibinfo {year}
  {2021})}\BibitemShut {NoStop}%
\bibitem [{\citenamefont {{LHCb collaboration}}(2020)}]{LHCb2020}%
  \BibitemOpen
  \bibfield  {author} {\bibinfo {author} {\bibnamefont {{LHCb
  collaboration}}},\ }\href {\doibase 10.1016/j.scib.2020.08.032} {\bibfield
  {journal} {\bibinfo  {journal} {Science Bulletin}\ }\textbf {\bibinfo
  {volume} {65}},\ \bibinfo {pages} {1983} (\bibinfo {year}
  {2020})}\BibitemShut {NoStop}%
\bibitem [{\citenamefont {Richard}(2020)}]{Richard2020}%
  \BibitemOpen
  \bibfield  {author} {\bibinfo {author} {\bibfnamefont {J.-M.}\ \bibnamefont
  {Richard}},\ }\href {\doibase 10.1016/j.scib.2020.08.020} {\bibfield
  {journal} {\bibinfo  {journal} {Sci. Bull.}\ }\textbf {\bibinfo {volume}
  {65}},\ \bibinfo {pages} {1954} (\bibinfo {year} {2020})}\BibitemShut
  {NoStop}%
\bibitem [{\citenamefont {Sonnenschein}\ and\ \citenamefont
  {Weissman}(2021)}]{Sonnenschein2020}%
  \BibitemOpen
  \bibfield  {author} {\bibinfo {author} {\bibfnamefont {J.}~\bibnamefont
  {Sonnenschein}}\ and\ \bibinfo {author} {\bibfnamefont {D.}~\bibnamefont
  {Weissman}},\ }\href {\doibase 10.1140/epjc/s10052-020-08818-7} {\bibfield
  {journal} {\bibinfo  {journal} {Eur. Phys. J. C}\ }\textbf {\bibinfo {volume}
  {81}},\ \bibinfo {pages} {25} (\bibinfo {year} {2021})}\BibitemShut {NoStop}%
\bibitem [{\citenamefont {Faustov}\ \emph {et~al.}(2021)\citenamefont
  {Faustov}, \citenamefont {Galkin},\ and\ \citenamefont
  {Savchenko}}]{Faustov2021}%
  \BibitemOpen
  \bibfield  {author} {\bibinfo {author} {\bibfnamefont {R.~N.}\ \bibnamefont
  {Faustov}}, \bibinfo {author} {\bibfnamefont {V.~O.}\ \bibnamefont {Galkin}},
  \ and\ \bibinfo {author} {\bibfnamefont {E.~M.}\ \bibnamefont {Savchenko}},\
  }\href {\doibase 10.3390/universe7040094} {\bibfield  {journal} {\bibinfo
  {journal} {Universe}\ }\textbf {\bibinfo {volume} {7}},\ \bibinfo {pages}
  {94} (\bibinfo {year} {2021})}\BibitemShut {NoStop}%
\bibitem [{\citenamefont {Deng}\ \emph {et~al.}(2021)\citenamefont {Deng},
  \citenamefont {Chen},\ and\ \citenamefont {Ping}}]{Deng2020}%
  \BibitemOpen
  \bibfield  {author} {\bibinfo {author} {\bibfnamefont {C.}~\bibnamefont
  {Deng}}, \bibinfo {author} {\bibfnamefont {H.}~\bibnamefont {Chen}}, \ and\
  \bibinfo {author} {\bibfnamefont {J.}~\bibnamefont {Ping}},\ }\href {\doibase
  10.1103/PhysRevD.103.014001} {\bibfield  {journal} {\bibinfo  {journal}
  {Phys. Rev. D}\ }\textbf {\bibinfo {volume} {103}},\ \bibinfo {pages}
  {014001} (\bibinfo {year} {2021})}\BibitemShut {NoStop}%
\bibitem [{\citenamefont {Guo}\ and\ \citenamefont {Oller}(2021)}]{Guo2020}%
  \BibitemOpen
  \bibfield  {author} {\bibinfo {author} {\bibfnamefont {Z.-H.}\ \bibnamefont
  {Guo}}\ and\ \bibinfo {author} {\bibfnamefont {J.~A.}\ \bibnamefont
  {Oller}},\ }\href {\doibase 10.1103/PhysRevD.103.034024} {\bibfield
  {journal} {\bibinfo  {journal} {Phys. Rev. D}\ }\textbf {\bibinfo {volume}
  {103}},\ \bibinfo {pages} {034024} (\bibinfo {year} {2021})}\BibitemShut
  {NoStop}%
\bibitem [{\citenamefont {Chen}\ \emph {et~al.}(2022)\citenamefont {Chen},
  \citenamefont {Chen}, \citenamefont {Liu}, \citenamefont {Liu},\ and\
  \citenamefont {Zhu}}]{Chen2022}%
  \BibitemOpen
  \bibfield  {author} {\bibinfo {author} {\bibfnamefont {H.-X.}\ \bibnamefont
  {Chen}}, \bibinfo {author} {\bibfnamefont {W.}~\bibnamefont {Chen}}, \bibinfo
  {author} {\bibfnamefont {X.}~\bibnamefont {Liu}}, \bibinfo {author}
  {\bibfnamefont {Y.-R.}\ \bibnamefont {Liu}}, \ and\ \bibinfo {author}
  {\bibfnamefont {S.-L.}\ \bibnamefont {Zhu}},\ }\href@noop {} {\  (\bibinfo
  {year} {2022})},\ \Eprint {http://arxiv.org/abs/2204.02649} {arXiv:2204.02649
  [hep-ph]} \BibitemShut {NoStop}%
\bibitem [{\citenamefont {Karliner}\ and\ \citenamefont
  {Rosner}(2020)}]{Karliner2020}%
  \BibitemOpen
  \bibfield  {author} {\bibinfo {author} {\bibfnamefont {M.}~\bibnamefont
  {Karliner}}\ and\ \bibinfo {author} {\bibfnamefont {J.~L.}\ \bibnamefont
  {Rosner}},\ }\href {\doibase 10.1103/PhysRevD.102.114039} {\bibfield
  {journal} {\bibinfo  {journal} {Phys. Rev. D}\ }\textbf {\bibinfo {volume}
  {102}},\ \bibinfo {pages} {114039} (\bibinfo {year} {2020})}\BibitemShut
  {NoStop}%
\bibitem [{\citenamefont {Debastiani}\ and\ \citenamefont
  {Navarra}(2019)}]{Debastiani2019}%
  \BibitemOpen
  \bibfield  {author} {\bibinfo {author} {\bibfnamefont {V.~R.}\ \bibnamefont
  {Debastiani}}\ and\ \bibinfo {author} {\bibfnamefont {F.~S.}\ \bibnamefont
  {Navarra}},\ }\href {\doibase 10.1088/1674-1137/43/1/013105} {\bibfield
  {journal} {\bibinfo  {journal} {Chin. Phys. C}\ }\textbf {\bibinfo {volume}
  {43}},\ \bibinfo {pages} {013105} (\bibinfo {year} {2019})}\BibitemShut
  {NoStop}%
\bibitem [{\citenamefont {Liu}\ \emph {et~al.}(2020)\citenamefont {Liu},
  \citenamefont {Liu}, \citenamefont {Zhong},\ and\ \citenamefont
  {Zhao}}]{Liu2020}%
  \BibitemOpen
  \bibfield  {author} {\bibinfo {author} {\bibfnamefont {M.-S.}\ \bibnamefont
  {Liu}}, \bibinfo {author} {\bibfnamefont {F.-X.}\ \bibnamefont {Liu}},
  \bibinfo {author} {\bibfnamefont {X.-H.}\ \bibnamefont {Zhong}}, \ and\
  \bibinfo {author} {\bibfnamefont {Q.}~\bibnamefont {Zhao}},\ }\href@noop {}
  {\enquote {\bibinfo {title} {{Full-heavy tetraquark states and their
  evidences in the LHCb di-$J/\psi$ spectrum}},}\ } (\bibinfo {year} {2020}),\
  \Eprint {http://arxiv.org/abs/2006.11952} {arXiv:2006.11952 [hep-ph]}
  \BibitemShut {NoStop}%
\bibitem [{\citenamefont {L\"u}\ \emph {et~al.}(2020)\citenamefont {L\"u},
  \citenamefont {Chen},\ and\ \citenamefont {Dong}}]{Lu2020}%
  \BibitemOpen
  \bibfield  {author} {\bibinfo {author} {\bibfnamefont {Q.-F.}\ \bibnamefont
  {L\"u}}, \bibinfo {author} {\bibfnamefont {D.-Y.}\ \bibnamefont {Chen}}, \
  and\ \bibinfo {author} {\bibfnamefont {Y.-B.}\ \bibnamefont {Dong}},\ }\href
  {\doibase 10.1140/epjc/s10052-020-08454-1} {\bibfield  {journal} {\bibinfo
  {journal} {Eur. Phys. J. C}\ }\textbf {\bibinfo {volume} {80}},\ \bibinfo
  {pages} {871} (\bibinfo {year} {2020})}\BibitemShut {NoStop}%
\bibitem [{\citenamefont {Wu}\ \emph {et~al.}(2018)\citenamefont {Wu},
  \citenamefont {Liu}, \citenamefont {Chen}, \citenamefont {Liu},\ and\
  \citenamefont {Zhu}}]{Wu2016}%
  \BibitemOpen
  \bibfield  {author} {\bibinfo {author} {\bibfnamefont {J.}~\bibnamefont
  {Wu}}, \bibinfo {author} {\bibfnamefont {Y.-R.}\ \bibnamefont {Liu}},
  \bibinfo {author} {\bibfnamefont {K.}~\bibnamefont {Chen}}, \bibinfo {author}
  {\bibfnamefont {X.}~\bibnamefont {Liu}}, \ and\ \bibinfo {author}
  {\bibfnamefont {S.-L.}\ \bibnamefont {Zhu}},\ }\href {\doibase
  10.1103/PhysRevD.97.094015} {\bibfield  {journal} {\bibinfo  {journal} {Phys.
  Rev. D}\ }\textbf {\bibinfo {volume} {97}},\ \bibinfo {pages} {094015}
  (\bibinfo {year} {2018})}\BibitemShut {NoStop}%
\bibitem [{\citenamefont {Bedolla}\ \emph {et~al.}(2020)\citenamefont
  {Bedolla}, \citenamefont {Ferretti}, \citenamefont {Roberts},\ and\
  \citenamefont {Santopinto}}]{Bedolla2019}%
  \BibitemOpen
  \bibfield  {author} {\bibinfo {author} {\bibfnamefont {M.~A.}\ \bibnamefont
  {Bedolla}}, \bibinfo {author} {\bibfnamefont {J.}~\bibnamefont {Ferretti}},
  \bibinfo {author} {\bibfnamefont {C.~D.}\ \bibnamefont {Roberts}}, \ and\
  \bibinfo {author} {\bibfnamefont {E.}~\bibnamefont {Santopinto}},\ }\href
  {\doibase 10.1140/epjc/s10052-020-08579-3} {\bibfield  {journal} {\bibinfo
  {journal} {Eur. Phys. J. C}\ }\textbf {\bibinfo {volume} {80}},\ \bibinfo
  {pages} {1004} (\bibinfo {year} {2020})}\BibitemShut {NoStop}%
\bibitem [{\citenamefont {Wang}\ \emph {et~al.}(2019)\citenamefont {Wang},
  \citenamefont {Meng},\ and\ \citenamefont {Zhu}}]{Wang2019}%
  \BibitemOpen
  \bibfield  {author} {\bibinfo {author} {\bibfnamefont {G.-J.}\ \bibnamefont
  {Wang}}, \bibinfo {author} {\bibfnamefont {L.}~\bibnamefont {Meng}}, \ and\
  \bibinfo {author} {\bibfnamefont {S.-L.}\ \bibnamefont {Zhu}},\ }\href
  {\doibase 10.1103/PhysRevD.100.096013} {\bibfield  {journal} {\bibinfo
  {journal} {Phys. Rev. D}\ }\textbf {\bibinfo {volume} {100}},\ \bibinfo
  {pages} {096013} (\bibinfo {year} {2019})}\BibitemShut {NoStop}%
\bibitem [{\citenamefont {Wan}\ and\ \citenamefont {Qiao}(2021)}]{Wan2020}%
  \BibitemOpen
  \bibfield  {author} {\bibinfo {author} {\bibfnamefont {B.-D.}\ \bibnamefont
  {Wan}}\ and\ \bibinfo {author} {\bibfnamefont {C.-F.}\ \bibnamefont {Qiao}},\
  }\href {\doibase 10.1016/j.physletb.2021.136339} {\bibfield  {journal}
  {\bibinfo  {journal} {Phys. Lett. B}\ }\textbf {\bibinfo {volume} {817}},\
  \bibinfo {pages} {136339} (\bibinfo {year} {2021})}\BibitemShut {NoStop}%
\bibitem [{\citenamefont {Liang}\ \emph {et~al.}(2021)\citenamefont {Liang},
  \citenamefont {Wu},\ and\ \citenamefont {Yao}}]{Liang2021}%
  \BibitemOpen
  \bibfield  {author} {\bibinfo {author} {\bibfnamefont {Z.-R.}\ \bibnamefont
  {Liang}}, \bibinfo {author} {\bibfnamefont {X.-Y.}\ \bibnamefont {Wu}}, \
  and\ \bibinfo {author} {\bibfnamefont {D.-L.}\ \bibnamefont {Yao}},\ }\href
  {\doibase 10.1103/PhysRevD.104.034034} {\bibfield  {journal} {\bibinfo
  {journal} {Phys. Rev. D}\ }\textbf {\bibinfo {volume} {104}},\ \bibinfo
  {pages} {034034} (\bibinfo {year} {2021})}\BibitemShut {NoStop}%
\bibitem [{\citenamefont {Li}\ \emph {et~al.}(2021)\citenamefont {Li},
  \citenamefont {Chang}, \citenamefont {Wang},\ and\ \citenamefont
  {Wang}}]{Li2021}%
  \BibitemOpen
  \bibfield  {author} {\bibinfo {author} {\bibfnamefont {Q.}~\bibnamefont
  {Li}}, \bibinfo {author} {\bibfnamefont {C.-H.}\ \bibnamefont {Chang}},
  \bibinfo {author} {\bibfnamefont {G.-L.}\ \bibnamefont {Wang}}, \ and\
  \bibinfo {author} {\bibfnamefont {T.}~\bibnamefont {Wang}},\ }\href {\doibase
  10.1103/PhysRevD.104.014018} {\bibfield  {journal} {\bibinfo  {journal}
  {Phys. Rev. D}\ }\textbf {\bibinfo {volume} {104}},\ \bibinfo {pages}
  {014018} (\bibinfo {year} {2021})}\BibitemShut {NoStop}%
\bibitem [{\citenamefont {Ke}\ \emph {et~al.}(2021)\citenamefont {Ke},
  \citenamefont {Han}, \citenamefont {Liu},\ and\ \citenamefont
  {Shi}}]{Ke2021}%
  \BibitemOpen
  \bibfield  {author} {\bibinfo {author} {\bibfnamefont {H.-W.}\ \bibnamefont
  {Ke}}, \bibinfo {author} {\bibfnamefont {X.}~\bibnamefont {Han}}, \bibinfo
  {author} {\bibfnamefont {X.-H.}\ \bibnamefont {Liu}}, \ and\ \bibinfo
  {author} {\bibfnamefont {Y.-L.}\ \bibnamefont {Shi}},\ }\href {\doibase
  10.1140/epjc/s10052-021-09229-y} {\bibfield  {journal} {\bibinfo  {journal}
  {Eur. Phys. J. C}\ }\textbf {\bibinfo {volume} {81}},\ \bibinfo {pages} {427}
  (\bibinfo {year} {2021})}\BibitemShut {NoStop}%
\bibitem [{\citenamefont {Weng}\ \emph {et~al.}(2021)\citenamefont {Weng},
  \citenamefont {Chen}, \citenamefont {Deng},\ and\ \citenamefont
  {Zhu}}]{Weng2020}%
  \BibitemOpen
  \bibfield  {author} {\bibinfo {author} {\bibfnamefont {X.-Z.}\ \bibnamefont
  {Weng}}, \bibinfo {author} {\bibfnamefont {X.-L.}\ \bibnamefont {Chen}},
  \bibinfo {author} {\bibfnamefont {W.-Z.}\ \bibnamefont {Deng}}, \ and\
  \bibinfo {author} {\bibfnamefont {S.-L.}\ \bibnamefont {Zhu}},\ }\href
  {\doibase 10.1103/PhysRevD.103.034001} {\bibfield  {journal} {\bibinfo
  {journal} {Phys. Rev. D}\ }\textbf {\bibinfo {volume} {103}},\ \bibinfo
  {pages} {034001} (\bibinfo {year} {2021})}\BibitemShut {NoStop}%
\bibitem [{\citenamefont {Zhu}(2021)}]{Zhu2020}%
  \BibitemOpen
  \bibfield  {author} {\bibinfo {author} {\bibfnamefont {R.}~\bibnamefont
  {Zhu}},\ }\href {\doibase 10.1016/j.nuclphysb.2021.115393} {\bibfield
  {journal} {\bibinfo  {journal} {Nucl. Phys. B}\ }\textbf {\bibinfo {volume}
  {966}},\ \bibinfo {pages} {115393} (\bibinfo {year} {2021})}\BibitemShut
  {NoStop}%
\bibitem [{\citenamefont {Pascalutsa}\ and\ \citenamefont
  {Vanderhaeghen}(2010)}]{Pascalutsa:2010sj}%
  \BibitemOpen
  \bibfield  {author} {\bibinfo {author} {\bibfnamefont {V.}~\bibnamefont
  {Pascalutsa}}\ and\ \bibinfo {author} {\bibfnamefont {M.}~\bibnamefont
  {Vanderhaeghen}},\ }\href {\doibase 10.1103/PhysRevLett.105.201603}
  {\bibfield  {journal} {\bibinfo  {journal} {Phys.\ Rev.\ Lett.}\ }\textbf
  {\bibinfo {volume} {105}},\ \bibinfo {pages} {201603} (\bibinfo {year}
  {2010})}\BibitemShut {NoStop}%
\bibitem [{\citenamefont {Pascalutsa}(2018)}]{Pascalutsa:2018ced}%
  \BibitemOpen
  \bibfield  {author} {\bibinfo {author} {\bibfnamefont {V.}~\bibnamefont
  {Pascalutsa}},\ }\href {\doibase 10.1088/978-1-6817-4919-8} {\emph {\bibinfo
  {title} {{Causality Rules}: {A light treatise on dispersion relations and sum
  rules}}}},\ IOP Concise Physics\ (\bibinfo  {publisher} {Morgan \& Claypool
  Publishers},\ \bibinfo {year} {2018})\BibitemShut {NoStop}%
\bibitem [{\citenamefont {Budnev}\ \emph {et~al.}(1975)\citenamefont {Budnev},
  \citenamefont {Ginzburg}, \citenamefont {Meledin},\ and\ \citenamefont
  {Serbo}}]{Budnev:1975poe}%
  \BibitemOpen
  \bibfield  {author} {\bibinfo {author} {\bibfnamefont {V.~M.}\ \bibnamefont
  {Budnev}}, \bibinfo {author} {\bibfnamefont {I.~F.}\ \bibnamefont
  {Ginzburg}}, \bibinfo {author} {\bibfnamefont {G.~V.}\ \bibnamefont
  {Meledin}}, \ and\ \bibinfo {author} {\bibfnamefont {V.~G.}\ \bibnamefont
  {Serbo}},\ }\href {\doibase 10.1016/0370-1573(75)90009-5} {\bibfield
  {journal} {\bibinfo  {journal} {Phys. Rept.}\ }\textbf {\bibinfo {volume}
  {15}},\ \bibinfo {pages} {181} (\bibinfo {year} {1975})}\BibitemShut
  {NoStop}%
\bibitem [{\citenamefont {Pascalutsa}\ \emph {et~al.}(2012)\citenamefont
  {Pascalutsa}, \citenamefont {Pauk},\ and\ \citenamefont
  {Vanderhaeghen}}]{Pascalutsa2012}%
  \BibitemOpen
  \bibfield  {author} {\bibinfo {author} {\bibfnamefont {V.}~\bibnamefont
  {Pascalutsa}}, \bibinfo {author} {\bibfnamefont {V.}~\bibnamefont {Pauk}}, \
  and\ \bibinfo {author} {\bibfnamefont {M.}~\bibnamefont {Vanderhaeghen}},\
  }\href {\doibase 10.1103/PhysRevD.85.116001} {\bibfield  {journal} {\bibinfo
  {journal} {Phys. Rev. D}\ }\textbf {\bibinfo {volume} {85}},\ \bibinfo
  {pages} {116001} (\bibinfo {year} {2012})}\BibitemShut {NoStop}%
\bibitem [{\citenamefont {Harland-Lang}\ \emph {et~al.}(2019)\citenamefont
  {Harland-Lang}, \citenamefont {Khoze},\ and\ \citenamefont
  {Ryskin}}]{HarlandLang2019}%
  \BibitemOpen
  \bibfield  {author} {\bibinfo {author} {\bibfnamefont {L.~A.}\ \bibnamefont
  {Harland-Lang}}, \bibinfo {author} {\bibfnamefont {V.~A.}\ \bibnamefont
  {Khoze}}, \ and\ \bibinfo {author} {\bibfnamefont {M.~G.}\ \bibnamefont
  {Ryskin}},\ }\href {\doibase 10.1140/epjc/s10052-018-6530-5} {\bibfield
  {journal} {\bibinfo  {journal} {The European Physical Journal C}\ }\textbf
  {\bibinfo {volume} {79}} (\bibinfo {year} {2019}),\
  10.1140/epjc/s10052-018-6530-5}\BibitemShut {NoStop}%
\bibitem [{\citenamefont {Harland-Lang}\ \emph {et~al.}(2016)\citenamefont
  {Harland-Lang}, \citenamefont {Khoze},\ and\ \citenamefont
  {Ryskin}}]{HarlandLang2016}%
  \BibitemOpen
  \bibfield  {author} {\bibinfo {author} {\bibfnamefont {L.~A.}\ \bibnamefont
  {Harland-Lang}}, \bibinfo {author} {\bibfnamefont {V.~A.}\ \bibnamefont
  {Khoze}}, \ and\ \bibinfo {author} {\bibfnamefont {M.~G.}\ \bibnamefont
  {Ryskin}},\ }\href {\doibase 10.1140/epjc/s10052-015-3832-8} {\bibfield
  {journal} {\bibinfo  {journal} {The European Physical Journal C}\ }\textbf
  {\bibinfo {volume} {76}} (\bibinfo {year} {2016}),\
  10.1140/epjc/s10052-015-3832-8}\BibitemShut {NoStop}%
\bibitem [{\citenamefont {Wang}\ \emph {et~al.}(2018)\citenamefont {Wang},
  \citenamefont {Sun}, \citenamefont {Liu},\ and\ \citenamefont
  {Matsuki}}]{Wang2018}%
  \BibitemOpen
  \bibfield  {author} {\bibinfo {author} {\bibfnamefont {J.~Z.}\ \bibnamefont
  {Wang}}, \bibinfo {author} {\bibfnamefont {Z.~F.}\ \bibnamefont {Sun}},
  \bibinfo {author} {\bibfnamefont {X.}~\bibnamefont {Liu}}, \ and\ \bibinfo
  {author} {\bibfnamefont {T.}~\bibnamefont {Matsuki}},\ }\href {\doibase
  10.1140/epjc/s10052-018-6372-1} {\bibfield  {journal} {\bibinfo  {journal}
  {European Physical Journal C}\ }\textbf {\bibinfo {volume} {78}},\ \bibinfo
  {pages} {1} (\bibinfo {year} {2018})}\BibitemShut {NoStop}%
\bibitem [{\citenamefont {Bern}\ \emph {et~al.}(2001)\citenamefont {Bern},
  \citenamefont {De~Freitas}, \citenamefont {Dixon}, \citenamefont
  {Ghinculov},\ and\ \citenamefont {Wong}}]{Bern:2001dg}%
  \BibitemOpen
  \bibfield  {author} {\bibinfo {author} {\bibfnamefont {Z.}~\bibnamefont
  {Bern}}, \bibinfo {author} {\bibfnamefont {A.}~\bibnamefont {De~Freitas}},
  \bibinfo {author} {\bibfnamefont {L.~J.}\ \bibnamefont {Dixon}}, \bibinfo
  {author} {\bibfnamefont {A.}~\bibnamefont {Ghinculov}}, \ and\ \bibinfo
  {author} {\bibfnamefont {H.~L.}\ \bibnamefont {Wong}},\ }\href {\doibase
  10.1088/1126-6708/2001/11/031} {\bibfield  {journal} {\bibinfo  {journal}
  {JHEP}\ }\textbf {\bibinfo {volume} {11}},\ \bibinfo {pages} {031} (\bibinfo
  {year} {2001})}\BibitemShut {NoStop}%
\bibitem [{\citenamefont {K\l{}usek-Gawenda}\ \emph {et~al.}(2016)\citenamefont
  {K\l{}usek-Gawenda}, \citenamefont {Sch\"afer},\ and\ \citenamefont
  {Szczurek}}]{Klusek-Gawenda:2016nuo}%
  \BibitemOpen
  \bibfield  {author} {\bibinfo {author} {\bibfnamefont {M.}~\bibnamefont
  {K\l{}usek-Gawenda}}, \bibinfo {author} {\bibfnamefont {W.}~\bibnamefont
  {Sch\"afer}}, \ and\ \bibinfo {author} {\bibfnamefont {A.}~\bibnamefont
  {Szczurek}},\ }\href {\doibase 10.1016/j.physletb.2016.08.059} {\bibfield
  {journal} {\bibinfo  {journal} {Phys. Lett. B}\ }\textbf {\bibinfo {volume}
  {761}},\ \bibinfo {pages} {399} (\bibinfo {year} {2016})}\BibitemShut
  {NoStop}%
\bibitem [{\citenamefont {Krintiras}\ \emph {et~al.}(2022)\citenamefont
  {Krintiras}, \citenamefont {Grabowska-Bold}, \citenamefont
  {K\l{}usek-Gawenda}, \citenamefont {Chapon}, \citenamefont {Chudasama},\ and\
  \citenamefont {Granier~de Cassagnac}}]{Krintiras:2022jxa}%
  \BibitemOpen
  \bibfield  {author} {\bibinfo {author} {\bibfnamefont {G.~K.}\ \bibnamefont
  {Krintiras}}, \bibinfo {author} {\bibfnamefont {I.}~\bibnamefont
  {Grabowska-Bold}}, \bibinfo {author} {\bibfnamefont {M.}~\bibnamefont
  {K\l{}usek-Gawenda}}, \bibinfo {author} {\bibfnamefont {E.}~\bibnamefont
  {Chapon}}, \bibinfo {author} {\bibfnamefont {R.}~\bibnamefont {Chudasama}}, \
  and\ \bibinfo {author} {\bibfnamefont {R.}~\bibnamefont {Granier~de
  Cassagnac}},\ }\href@noop {} {\  (\bibinfo {year} {2022})},\ \Eprint
  {http://arxiv.org/abs/2204.02845} {arXiv:2204.02845 [hep-ph]} \BibitemShut
  {NoStop}%
\bibitem [{\citenamefont {Barnea}\ \emph {et~al.}(2006)\citenamefont {Barnea},
  \citenamefont {Vijande},\ and\ \citenamefont {Valcarce}}]{Barnea2006}%
  \BibitemOpen
  \bibfield  {author} {\bibinfo {author} {\bibfnamefont {N.}~\bibnamefont
  {Barnea}}, \bibinfo {author} {\bibfnamefont {J.}~\bibnamefont {Vijande}}, \
  and\ \bibinfo {author} {\bibfnamefont {A.}~\bibnamefont {Valcarce}},\ }\href
  {\doibase 10.1103/PhysRevD.73.054004} {\bibfield  {journal} {\bibinfo
  {journal} {Phys. Rev. D}\ }\textbf {\bibinfo {volume} {73}},\ \bibinfo
  {pages} {054004} (\bibinfo {year} {2006})}\BibitemShut {NoStop}%
\bibitem [{\citenamefont {Berezhnoy}\ \emph {et~al.}(2012)\citenamefont
  {Berezhnoy}, \citenamefont {Luchinsky},\ and\ \citenamefont
  {Novoselov}}]{Berezhnoy2011}%
  \BibitemOpen
  \bibfield  {author} {\bibinfo {author} {\bibfnamefont {A.~V.}\ \bibnamefont
  {Berezhnoy}}, \bibinfo {author} {\bibfnamefont {A.~V.}\ \bibnamefont
  {Luchinsky}}, \ and\ \bibinfo {author} {\bibfnamefont {A.~A.}\ \bibnamefont
  {Novoselov}},\ }\href {\doibase 10.1103/PhysRevD.86.034004} {\bibfield
  {journal} {\bibinfo  {journal} {Phys. Rev. D}\ }\textbf {\bibinfo {volume}
  {86}},\ \bibinfo {pages} {034004} (\bibinfo {year} {2012})}\BibitemShut
  {NoStop}%
\bibitem [{\citenamefont {Karliner}\ \emph {et~al.}(2017)\citenamefont
  {Karliner}, \citenamefont {Nussinov},\ and\ \citenamefont
  {Rosner}}]{Karliner2016}%
  \BibitemOpen
  \bibfield  {author} {\bibinfo {author} {\bibfnamefont {M.}~\bibnamefont
  {Karliner}}, \bibinfo {author} {\bibfnamefont {S.}~\bibnamefont {Nussinov}},
  \ and\ \bibinfo {author} {\bibfnamefont {J.~L.}\ \bibnamefont {Rosner}},\
  }\href {\doibase 10.1103/PhysRevD.95.034011} {\bibfield  {journal} {\bibinfo
  {journal} {Phys. Rev. D}\ }\textbf {\bibinfo {volume} {95}},\ \bibinfo
  {pages} {034011} (\bibinfo {year} {2017})}\BibitemShut {NoStop}%
\bibitem [{\citenamefont {Wang}(2017)}]{Wang2017}%
  \BibitemOpen
  \bibfield  {author} {\bibinfo {author} {\bibfnamefont {Z.-G.}\ \bibnamefont
  {Wang}},\ }\href {\doibase 10.1140/epjc/s10052-017-4997-0} {\bibfield
  {journal} {\bibinfo  {journal} {Eur. Phys. J. C}\ }\textbf {\bibinfo {volume}
  {77}},\ \bibinfo {pages} {432} (\bibinfo {year} {2017})}\BibitemShut
  {NoStop}%
\bibitem [{\citenamefont {Liu}\ \emph {et~al.}(2019)\citenamefont {Liu},
  \citenamefont {L\"u}, \citenamefont {Zhong},\ and\ \citenamefont
  {Zhao}}]{Liu2019}%
  \BibitemOpen
  \bibfield  {author} {\bibinfo {author} {\bibfnamefont {M.-S.}\ \bibnamefont
  {Liu}}, \bibinfo {author} {\bibfnamefont {Q.-F.}\ \bibnamefont {L\"u}},
  \bibinfo {author} {\bibfnamefont {X.-H.}\ \bibnamefont {Zhong}}, \ and\
  \bibinfo {author} {\bibfnamefont {Q.}~\bibnamefont {Zhao}},\ }\href {\doibase
  10.1103/PhysRevD.100.016006} {\bibfield  {journal} {\bibinfo  {journal}
  {Phys. Rev. D}\ }\textbf {\bibinfo {volume} {100}},\ \bibinfo {pages}
  {016006} (\bibinfo {year} {2019})}\BibitemShut {NoStop}%
\bibitem [{\citenamefont {Lundhammar}\ and\ \citenamefont
  {Ohlsson}(2020)}]{Lundhammar2020}%
  \BibitemOpen
  \bibfield  {author} {\bibinfo {author} {\bibfnamefont {P.}~\bibnamefont
  {Lundhammar}}\ and\ \bibinfo {author} {\bibfnamefont {T.}~\bibnamefont
  {Ohlsson}},\ }\href {\doibase 10.1103/PhysRevD.102.054018} {\bibfield
  {journal} {\bibinfo  {journal} {Phys. Rev. D}\ }\textbf {\bibinfo {volume}
  {102}},\ \bibinfo {pages} {054018} (\bibinfo {year} {2020})}\BibitemShut
  {NoStop}%
\bibitem [{\citenamefont {Barger}\ and\ \citenamefont
  {Phillips}(1975)}]{Barger1975}%
  \BibitemOpen
  \bibfield  {author} {\bibinfo {author} {\bibfnamefont {V.}~\bibnamefont
  {Barger}}\ and\ \bibinfo {author} {\bibfnamefont {R.}~\bibnamefont
  {Phillips}},\ }\href {\doibase 10.1016/0370-2693(75)90582-1} {\bibfield
  {journal} {\bibinfo  {journal} {Physics Letters B}\ }\textbf {\bibinfo
  {volume} {58}},\ \bibinfo {pages} {433} (\bibinfo {year} {1975})}\BibitemShut
  {NoStop}%
\bibitem [{\citenamefont {Redlich}\ \emph {et~al.}(2000)\citenamefont
  {Redlich}, \citenamefont {Satz},\ and\ \citenamefont
  {Zinovjev}}]{Redlich2000}%
  \BibitemOpen
  \bibfield  {author} {\bibinfo {author} {\bibfnamefont {K.}~\bibnamefont
  {Redlich}}, \bibinfo {author} {\bibfnamefont {H.}~\bibnamefont {Satz}}, \
  and\ \bibinfo {author} {\bibfnamefont {G.~M.}\ \bibnamefont {Zinovjev}},\
  }\href {\doibase 10.1007/s100520000488} {\bibfield  {journal} {\bibinfo
  {journal} {Eur. Phys. J. C}\ }\textbf {\bibinfo {volume} {17}},\ \bibinfo
  {pages} {461} (\bibinfo {year} {2000})}\BibitemShut {NoStop}%
\bibitem [{\citenamefont {Zyla}\ \emph {et~al.}(2020)\citenamefont {Zyla} \emph
  {et~al.}}]{PDG2020}%
  \BibitemOpen
  \bibfield  {author} {\bibinfo {author} {\bibfnamefont {P.~A.}\ \bibnamefont
  {Zyla}} \emph {et~al.} (\bibinfo {collaboration} {Particle Data Group}),\
  }\href {\doibase 10.1093/ptep/ptaa104} {\bibfield  {journal} {\bibinfo
  {journal} {PTEP}\ }\textbf {\bibinfo {volume} {2020}},\ \bibinfo {pages}
  {083C01} (\bibinfo {year} {2020})}\BibitemShut {NoStop}%
\bibitem [{\citenamefont {Esposito}\ \emph {et~al.}(2021)\citenamefont
  {Esposito}, \citenamefont {Manzari}, \citenamefont {Pilloni},\ and\
  \citenamefont {Polosa}}]{Esposito2021}%
  \BibitemOpen
  \bibfield  {author} {\bibinfo {author} {\bibfnamefont {A.}~\bibnamefont
  {Esposito}}, \bibinfo {author} {\bibfnamefont {C.~A.}\ \bibnamefont
  {Manzari}}, \bibinfo {author} {\bibfnamefont {A.}~\bibnamefont {Pilloni}}, \
  and\ \bibinfo {author} {\bibfnamefont {A.~D.}\ \bibnamefont {Polosa}},\
  }\href {\doibase 10.1103/PhysRevD.104.114029} {\bibfield  {journal} {\bibinfo
   {journal} {Phys. Rev. D}\ }\textbf {\bibinfo {volume} {104}},\ \bibinfo
  {pages} {114029} (\bibinfo {year} {2021})}\BibitemShut {NoStop}%
\end{thebibliography}%

\end{document}